\documentclass[aps,floatfix,10pt,amssymb,amsfont,amsmath,superscriptaddress,float,nofootinbib,longbibliography,nobalancelastpage,twocolumn]{revtex4-2}
\usepackage{graphicx} 
\usepackage{braket}
\usepackage{hyperref}
\usepackage{amsmath}
\usepackage{amssymb}
\usepackage{yquant}
\hypersetup{hidelinks}

\begin{document}
\author{Oliver O'Brien}
\affiliation{Department of Applied Mathematics and Theoretical Physics, University of Cambridge,\\ Wilberforce Road, Cambridge, CB3 0WA, United Kingdom}
\title{Solving a non-abelian hidden subgroup problem on a trapped-ion\\ quantum computer: the power of the measurement basis}
\date{November 2026}
\begin{abstract}
The hidden subgroup problem (HSP) unifies many of the best-known quantum speed-ups. For abelian groups it is solved efficiently by weak Fourier sampling, but for non-abelian groups, which include the cases relevant to graph isomorphism and lattice-based cryptography, the right measurement basis is crucial and often unknown. We experimentally study the HSP over the $q$-hedral groups $\mathbb{Z}_q \ltimes \mathbb{Z}_p$, which interpolate between the dihedral and affine groups and mark a boundary between instances that are efficiently solvable and those that are only information-theoretically solvable. On the IonQ Forte trapped-ion quantum computer, with $p \le 7$, we implement weak Fourier sampling, strong Fourier sampling in an adapted basis, strong Fourier sampling of adjacent pairs, and forgetful abelian sampling. We demonstrate that the theoretical predictions of Moore, Rockmore, Russell and Schulman \cite{moore2007power} survive on a noisy device: weak sampling reveals nothing about the hidden subgroup, the adapted basis succeeds for large $q$, adjacent pairs succeed for small $q$, and forgetful sampling saturates below unit success. For $p=7$, $q=6$, the hardware exceeds the success probability of an exhaustively optimised classical algorithm for up to three oracle queries.
\end{abstract}
\maketitle
\section{Introduction}
Possibly the most famous future application of quantum computers is efficiently factoring numbers using Shor's Algorithm \cite{shor1999polynomial}, which could in principle break the RSA cryptography \cite{rivest1978method} that secures much of the world’s internet traffic. Factoring can be classically reduced to period finding \cite{shor1999polynomial}, and while no polynomial-time classical algorithm for factoring is known, a quantum computer can find periods efficiently using the quantum Fourier transform. Period finding is one instance of the abelian Hidden Subgroup Problem (HSP) \cite{kitaev1995quantum}, which can be solved efficiently on a quantum computer for any finite abelian group, whereas classical algorithms typically require exponentially many queries.

Many interesting problems can similarly be reduced to the non-abelian HSP, including graph isomorphism (symmetric group) \cite{beals1997quantum, hallgren2010limitations} and the unique shortest vector problem (dihedral group) \cite{regev2004quantum}, which is closely related to the problems underpinning lattice-based post-quantum cryptography. However, unlike the abelian case, no general efficient quantum algorithm is known for the non-abelian HSP. Despite this, some instances are efficiently solvable using carefully tailored quantum algorithms \cite{hallgren2000normal, bacon2005optimal, moore2007power, grigni2001quantum, ivanyos2001efficient, friedl2014hidden, morales2026hidden, Inui_2007}.

In this work we demonstrate the first experimental implementation of solving a non-abelian HSP on a quantum device. We explore the particularly interesting q-hedral family of non-abelian groups which delineate a boundary of efficiently solvable problems and those only solvable in an information theoretical sense. This family was first studied in this context by Moore, Rockmore, Russell and Schulman (MRRS) \cite{moore2002hidden, moore2007power}, and our work can be viewed as an experimental demonstration that their theoretical results survive on a noisy quantum device.

MRRS distinguish between \emph{fully reconstructible} HSP over group $G$, which can be solved using quantum circuits of size polynomial in $\log|G|$ followed by polynomial-time classical post-processing, and \emph{information-theoretically reconstructible} problems that are solvable with polynomial sized quantum circuits but the classical post-processing may be arbitrary. HSP over the dihedral group ($D_p$) is information-theoretically reconstructible as $O(\log p)$ quantum queries provide enough information to determine the hidden subgroup, but the best known classical decoding algorithm takes time exponential in $\log p$ \cite{ettinger1999quantum}\footnote{Kuperberg’s quantum algorithm also solves the dihedral HSP in subexponential time $2^{O(\sqrt{\log p})}$ \cite{kuperberg2005subexponential}.}. In this work we evaluate how carefully choosing the measurement basis is crucial to ensuring both full and information-theoretic reconstructibility.

The $q$-hedral groups interpolate between the dihedral group ($q=2$) and the affine group ($q=p-1$) for each prime $p$. The main results of MRRS are that the $q$-hedral HSP is fully reconstructible when $q = p/\operatorname{polylog} p$, using strong Fourier sampling in an adapted basis, and information-theoretically reconstructible for all $q$, using strong Fourier sampling of adjacent pairs. They also show that forgetful abelian sampling~\cite{ettinger1999quantum}, which discards the non-abelian structure of the group, does not suffice.

In this work we implement each of these measurement strategies end-to-end on the IonQ Forte trapped-ion quantum computer for $p \in \{3,5,7\}$ and a range of $q$ spanning the dihedral and affine extremes. Our contributions are as follows.
\begin{enumerate}
    \item We introduce a small modification to the adjacent-pairs measurement, replacing $H$ with $HT$, which removes an ambiguity between $b$ and $-b$ present in the original scheme.
    \item We construct hardware-efficient circuits for each component: numerically synthesised $QFT_p$ circuits, an oracle implemented by Fourier conjugation, and a Walsh-series synthesis of the diagonal gates required by the exponent encoding, each using substantially fewer two-qubit gates than standard compiler synthesis.
    \item We observe the signatures predicted by theory: weak Fourier sampling carries no information about the hidden subgroup; adapted-basis sampling succeeds with probability $(p-1)q/p^2$; adjacent-pairs sampling succeeds for small $q$, where the adapted basis fails; and forgetful abelian sampling plateaus at success probability $(p+q-1)/(pq)$ because it cannot distinguish $b$ within an orbit.
    \item We compare against an exhaustively optimised, derandomised classical birthday algorithm and find that for $p=7$, $q=6$ the noisy hardware achieves a higher success probability than the optimal classical strategy for up to three queries.
\end{enumerate}

At these sizes, the comparison concerns query complexity rather than computational speed-up. As in previous demonstrations of the abelian HSP~\cite{singkanipa2025demonstration}, a third party must act as ``compiler'', since the secret $b$ is compiled into the oracle circuit; knowledge of $b$ is used only to construct the oracle, never in decoding, and the oracle is treated as a black box when counting queries. In addition,  circuit elements acting only on registers that are discarded (the final $QFT_p^\dagger$ and permutation in the oracle, and the controlled uncomputation in the non-abelian QFT) are omitted, as detailed in Appendix~\ref{section:circuits}. To our knowledge, this is nonetheless the first experimental demonstration of strong Fourier sampling and of an end-to-end non-abelian hidden subgroup algorithm.

The remainder of this paper is organised as follows. Section~\ref{section:theory} reviews the $q$-hedral HSP and derives the measurement statistics of each sampling method. Section~\ref{section:results} presents the experimental results, and Section~\ref{section:discussion} discusses their implications. Appendix~\ref{section:circuits} describes our circuit constructions and error mitigation.

\section{q-Hedral Hidden Subgroup Problem}
\label{section:theory}
When solving the Hidden Subgroup Problem we first prepare a register with a uniform superposition over all of the elements of the group $G$. In our case we are considering the $q$-hedral group $G= \mathbb{Z}_q \ltimes \mathbb{Z}_p$ for a given prime $p$ and divisor $q$ of $p-1$. The group is composed of elements $(x,\alpha^t)$ with $x \in \mathbb{Z}_p$, $t \in \mathbb{Z}_q$, $\operatorname{ord}\alpha = q$ and the group action is given by $(x,\alpha^t)(x',\alpha^{t'}) = (x+\alpha^tx', \alpha^{t+t'})$. Therefore, we begin by preparing the initial state:
\begin{equation}
    \frac{1}{\sqrt{qp}}\sum_{x = 0}^{p-1}\sum_{t=0}^{q-1} \ket{x}\ket{\alpha^t}
    \label{eq:initialstate}
\end{equation}
We then act with an oracle invariant under the action of our hidden subgroup $H_b = \{(1-\alpha^t)b, \alpha^t) \>\forall\> t \in \mathbb{Z}_q\}$. In our instance the oracle we choose is:
\begin{equation}
    f_b(x,t) = \pi(x + \alpha^t b)
\end{equation}
where we choose a public $\alpha$ such that $\operatorname{ord}\alpha = q$, and $\pi$ is some random secret permutation. Our aim is to learn the secret value of $b$. After acting with our oracle we have the following state:
\begin{equation}
        \frac{1}{\sqrt{qp}}\sum_{x = 0}^{p-1}\sum_{t=0}^{q-1} \ket{x}\ket{\alpha^t}\ket{f_b(x,\alpha^t)}
\end{equation}
We then measure the register containing the output of the oracle getting measurement result $c$. As the oracle is invariant under the action of the subgroup $H_b$ we are now restricted to a superposition over the coset $cH_b$. Therefore, our post-measurement state is given by:
\begin{equation}
    \frac{1}{\sqrt{q}} \sum_{t=0}^{q-1}\ket{\pi^{-1}(c) - \alpha^tb}\ket{\alpha^t}
\end{equation}
The next step is to perform the QFT over the full group $G$. As the group is non-abelian the form of this QFT is more complex than usual. As defined by \cite{hallgren2000normal} the QFT of this coset state is given by:
\begin{equation}
    \sum_{\rho \in \tilde G} \frac{1}{q}\sqrt{\frac{d_{\rho}}{p}}\sum_{i,j}\sum_{t=0}^{q-1} \rho((\pi^{-1}(c) - \alpha^tb,\alpha^t))_{i,j} \ket{\rho,i,j}
    \label{eq:qftofcoset}
\end{equation}
where $\tilde G$ is the set of irreducible representations of $G$. In our instance $G$ has $q$ one-dimensional irreps given by:
\begin{equation}
    \sigma_l((x,t)) = \omega^{lt}_q
\end{equation}
for $l \in \mathbb{Z}_q$ and $(p-1)/q$ distinct $q$-dimensional irreps given by:
\begin{equation}
    \rho_l((x,t))_{j,k} = \begin{cases}
        \omega_p^{l \alpha^j x} & k = j + t \mod q\\
        0 & \text{else}
    \end{cases}
\end{equation}
for $l \in \mathbb{Z}_p^*/\mathbb{Z}_q$. Therefore, we can rewrite \eqref{eq:qftofcoset} as:
\begin{multline}
     \frac{1}{q}\sqrt{\frac{1}{p}}\sum_{t=0}^{q-1}  \sum_{l=0}^{q-1}\omega^{lt}_q\ket{\sigma_l,0,0} +  \\\sqrt{\frac{1}{pq}}\sum_{t=0}^{q-1} \sum_{l =0}^{\frac{p-1}{q}-1}\sum_{j=0}^{q-1} \omega_p^{g^{l+1}\alpha^j x}\ket{\rho_l,j,j+t} 
\end{multline}
where $g$ is a primitive root modulo $p$ and $x = \pi^{-1}(c) - \alpha^tb$. It is clear that $\alpha = g^\frac{p-1}{q}$ as $\alpha^q = g^{p-1} = 1 \mod p$. Every $l  \in \mathbb{Z}_{\frac{p-1}{q}}$ and $j  \in \mathbb{Z}_q$ gives a unique $k=g^{l+1}\alpha^j \mod p$ with $k \in \mathbb{Z}_p^*$, and hence we can perform the unique relabeling of the $(p-1)q$ basis states $\ket{\rho_l,j,j+t}$ as $\ket{g^{l+1}\alpha^j,\alpha^t}$. Therefore, we have:
\begin{multline}
     \frac{1}{q}\sqrt{\frac{1}{p}}\sum_{t=0}^{q-1}  \sum_{l=0}^{q-1}\omega^{lt}_q\ket{\sigma_l,0,0} + \\ \sqrt{\frac{1}{pq}}\sum_{t=0}^{q-1} \sum_{l =0}^{\frac{p-1}{q}-1}\sum_{j=0}^{q-1} \omega_p^{g^{l+1} \alpha^j x}\ket{g^{l+1}\alpha^j,\alpha^t}
\end{multline}
as $g^{l+1} \alpha^j$ generates every $k \in \mathbb{Z}_p^*$ exactly once when we sum over $l \in \mathbb{Z}_{\frac{p-1}{q}}$ and $j \in \mathbb{Z}_q$, and hence we can replace the summations with one over $k \in \mathbb{Z}_p^*$:
\begin{equation}
     \frac{1}{q}\sqrt{\frac{1}{p}}\sum_{t=0}^{q-1}  \sum_{l=0}^{q-1}\omega^{lt}_q\ket{\sigma_l,0,0} +  \sqrt{\frac{1}{pq}}\sum_{t=0}^{q-1} \sum_{k=1}^{p-1} \omega_p^{k x}\ket{k}\ket{\alpha^t}
\end{equation}
Now we can perform a unique relabeling of the first term from $\ket{\sigma_l,0,0}$ to $\ket{0}\ket{l}$ giving:
\begin{multline}
     \frac{1}{q}\sqrt{\frac{1}{p}}\sum_{t=0}^{q-1}  \sum_{l=0}^{q-1}\omega^{lt}_q\ket{0}\ket{l} + \\ \sqrt{\frac{1}{pq}}\sum_{t=0}^{q-1} \sum_{k=1}^{p-1} \omega_p^{k (\pi^{-1}(c) - \alpha^tb)}\ket{k}\ket{\alpha^t}
\end{multline}
where we have substituted back in our expression for $x$. By taking the summation over $t$ in the first term we have:
\begin{equation}
     \sqrt{\frac{1}{p}} \ket{0}\ket{0} +  \sqrt{\frac{1}{pq}}\sum_{t=0}^{q-1} \sum_{k=1}^{p-1} \omega_p^{k (\pi^{-1}(c) - \alpha^tb)}\ket{k}\ket{\alpha^t}
     \label{eq:finalqftstate}
\end{equation}
We have now made a choice of basis that ensures the state after applying QFT has the same tensor product structure as before. 

After applying the Quantum Fourier Transform the next step of the standard Hidden Subgroup Problem algorithm is to measure the state. This measurement reveals an irrep label that the coset state transforms under. For abelian groups we can stop here as the distribution of irrep labels unique identifies the subgroup of the abelian group. However, for non-abelian groups such as ours many subgroups can share the same irrep label. Indeed measuring just the irrep label is categorised in the literature as ``weak fourier sampling" and is known to be sufficient to identify normal subgroups but insufficient for any non-normal subgroups. Therefore, the standard approach to solving HSP for non-abelian groups is to employ the ``strong fourier sampling" method. Here we still measure the irrep label, but do not simply discard the remaining state. Instead by employing a \emph{careful} choice of measurement basis we are able to extract information about the hidden subgroup. This approach is not always successful and has been demonstrated to be insufficient for the symmetric group \cite{moore2008symmetric}.

The optimal choice of measurement basis for ``strong fourier sampling" is not easily derived for a general non-abelian group. However, for the q-hedral group we consider there has been extensive research into different methods for performing ``strong fourier sampling". In this work we demonstrate practical implementations of a range of sampling methods and compare their performance on an existing quantum device. To our knowledge this is the first experimental implementation of ``strong fourier sampling" and the first experimental demonstration of end-to-end non-abelian HSP. Below we will discuss each of the sampling techniques we will demonstrate.

\subsection{Strong sampling in adapted basis}
After measuring the irrep label $k \neq 0$ (with probability $1-\frac{1}{p}$) we have the following post-measurement state:
\begin{equation}
    \sqrt{\frac{1}{q}}\sum_{t=0}^{q-1} \omega_p^{k (\pi^{-1}(c) - \alpha^tb)}\ket{\alpha^t}
\end{equation}
If we simply perform abelian $QFT_p$ on the remaining register to get:
\begin{equation}
    \sqrt{\frac{1}{qp}}\sum_{t=0}^{q-1}\sum_{l=0}^{p-1} \omega_p^{k \pi^{-1}(c) - \alpha^tbk + \alpha^t l}\ket{l}
\end{equation}
We then measure the register which gives $l$ with probability:
\begin{equation}
    \frac{1}{qp}\left|\sum_{t=0}^{q-1} \omega_p^{ \alpha^t(l-kb)}\right|^2
\end{equation}
Therefore, the outcome $l=kb \mod p$ occurs with probability $\frac{q}{p}$. Hence, as we know $k$ we are able to solve the Hidden Subgroup Problem by finding $b$ with probability $(1-\frac{1}{p})\frac{q}{p}$. Therefore, in the regime where $q = p/\text{polylog}(p)$ the Hidden Subgroup Problem on $\mathbb{Z}_q \ltimes \mathbb{Z}_p$ is fully reconstructible (solvable with $O(\text{polylog} (p))$ quantum samples by a quantum circuit of size $O(\text{polylog} (p))$).  However, in the regime of small $q$ this approach breaks down. For example, for constant $q$ we would require $\Omega(p)$ samples to determine $b$ using a mode based decoder which is less efficient than the $O(\sqrt{p})$ classical collision approach. Therefore, in this regime we use a different method of strong sampling described below.

\subsection{Strong sampling of adjacent pairs}
\label{section:pairs}

To implement this sampling technique on existing quantum devices we use a different encoding of the group to the previous method. Instead of storing $\alpha^t$ in the second register, we will simply store the exponent $t$. The only impact of this is upon the quantum circuit for applying the oracle and the state preparation technique. In this encoding the post-measurement state we obtain with probability $1-\frac{1}{p}$ is:
\begin{equation}
    \sqrt{\frac{1}{q}}\sum_{t=0}^{q-1} \omega_p^{k (\pi^{-1}(c) - \alpha^tb)}\ket{t}
\end{equation}
We proceed by applying $HT$ to the qubit storing the least significant bit of $t$, which gives the following state:
\begin{equation}
    \omega_p^{k \pi^{-1}(c)}\sqrt{\frac{1}{q}}\sum_{t=0}^{q/2-1} \omega_p^{- \alpha^{2t}bk}\ket{2t}\ket{+}+ e^{i\frac{\pi}{4}}\omega_p^{ - \alpha^{2t+1}bk}\ket{2t}\ket{-}
\end{equation}
\begin{multline}
    \omega_p^{k \pi^{-1}(c)}\sqrt{\frac{2}{q}}\sum_{t=0}^{q/2-1} \\e^{i\frac{\pi}{8}}\omega_p^{ -\frac{1}{2}(1+\alpha)\alpha^{2t}bk}\cos\left(\frac{\pi}{p}(1-\alpha)\alpha^{2t}bk+ \frac{\pi}{8}\right) \ket{2t}\ket{0}-\\ 
    ie^{i\frac{\pi}{8}}\omega_p^{ -\frac{1}{2}(1+\alpha)\alpha^{2t}bk}\sin \left( \frac{\pi}{p}(1-\alpha)\alpha^{2t}bk + \frac{\pi}{8}\right)\ket{2t}\ket{1}
\end{multline}
for even $q$. Therefore, upon measuring all the qubits we have that the expectation value of the least significant bit varies as $\sin \left( \frac{m\pi}{p}b + \frac{\pi}{8}\right)^2$ in $m=(1-\alpha)\alpha^{2t}k \mod p$. The original MRRS paper instead applied just $H$ leading to a $\sin \left( \frac{m\pi}{p}b\right)^2$ signal. They state that the total variation distance of this signal between any two distinct $b$ is bounded below by a constant, so in theory we can distinguish between $p$ different $b$ with only $O(\log p)$ samples for all $q$. However, under their signal it is impossible to distinguish $b$ from $p-b$ as:
\begin{equation}
    \sin \left( \frac{m\pi}{p}(p-b)\right)^2 =\sin \left(\frac{m\pi}{p}b\right)^2
\end{equation}
The total-variation bound of \cite{moore2007power} therefore holds only for pairs $b' \neq \pm b$. Our slight modification of applying $HT$ alleviates this and ensures that the Hidden Subgroup Problem on $\mathbb{Z}_q \ltimes \mathbb{Z}_p$ is information-theoretic reconstructible for all $q$. However, there is no known $O(\text{polylog } p)$ classical decoder, so it is not shown to be fully reconstructible for all $q$.

We can however decode the signal in $O(p \log p)$ time using FFT and hence demonstrate an oracle query advantage over a purely classical approach.

\subsection{Forgetful Abelian Method}
\label{section:forgetful}
As discussed in MRRS it is sometimes possible to solve HSP for a non-abelian group by erasing its non-abelian structure and performing an abelian Fourier transform instead. In our case this would correspond to treating the group instead as the direct product $\mathbb{Z}_q \times \mathbb{Z}_p$ with the corresponding abelian fourier transform given by $QFT_p \otimes QFT_q$. We can view this in within our existing framework as a choice of basis for strong fourier sampling with the exponent encoding described in the previous section.

Starting from the post-measurement state for any $k$ (including $k=0$):
\begin{equation}
    \sqrt{\frac{1}{q}}\sum_{t=0}^{q-1} \omega_p^{k (\pi^{-1}(c) - \alpha^tb)}\ket{t}
\end{equation}
we apply $QFT_q$ to get:
\begin{equation}
        \frac{1}{q}\sum_{l,t=0}^{q-1} \omega_p^{k (\pi^{-1}(c) - \alpha^tb)}\omega_q^{lt}\ket{l}
\end{equation}
Therefore, the probability of measuring irrep labels $k$ and $l$ is given by:
\begin{equation}
    P_b(k,l) = \frac{1}{pq^2}\left| \sum_{t=0}^{q-1} \omega_p^{ - \alpha^tbk}\omega_q^{lt}\right|^2
\end{equation}
Taking $b'=\alpha^m b$ for $m \in \mathbb{Z}_q$ gives:
\begin{align}
     P_{b'}(k,l) &= \frac{1}{pq^2}\left| \sum_{t=0}^{q-1} \omega_p^{ - \alpha^{t+m}bk}\omega_q^{lt}\right|^2 \\
     &= \frac{1}{pq^2}\left| \sum_{t'=0}^{q-1} \omega_p^{ - \alpha^{t'}bk}\omega_q^{l(t'-m)}\right|^2\\
     &= \frac{1}{pq^2}\left| \omega_q^{-lm}\sum_{t'=0}^{q-1} \omega_p^{ - \alpha^{t'}bk}\omega_q^{lt'}\right|^2\\
     &= P_b(k,l)
\end{align}
Therefore, this forgetful abelian sampling method cannot distinguish $b$ from $q-1$ other $b$ for $b\neq 0$. A decoder can therefore only identify the orbit of $b$, as the average number of elements in a orbit is $\frac{pq}{p+q-1}$ the success probability is capped at $\frac{p+q-1}{pq}$ regardless of the number of samples collected. This is evidence that not only is the choice of strong fourier sampling basis important for efficient decoding, but also effects the information theoretic reconstruction of the subgroup.

\section{Results}
\label{section:results}

We demonstrate the theoretical predictions of Section~\ref{section:theory} by running the quantum circuits described in Appendix~\ref{section:circuits} (Figs.~\ref{fig:circuit-adapted} and \ref{fig:circuit-exponent}) on the IonQ Forte quantum device \cite{chen2024benchmarking}. We restrict to primes $p=3,5,7$ as larger primes required larger circuits and subsequently too much hardware noise to observe a clear signal.  

\paragraph*{Experimental Details} All data were taken on the IonQ Forte trapped-ion computer between 9 and 12 October 2026. We ran 62 circuits (every $p \leq 7$, $q$ and secret $b$ shown in the figures) at 2,048 shots each, plus gate-folded (fold = 3) copies of 13 circuits, for 75 jobs and 153,600 shots in total. Circuits were compiled offline to Forte's native gates (GPi, GPi2 and arbitrary-angle ZZ) and submitted as native gate lists, so no further compilation removed the folded gates. Every job used IonQ's debiasing, which runs a job as 32 variants of 64 shots over different qubit assignments and combines their counts. We refer to these debiased counts, without further mitigation, as ``raw". 

\begin{figure}[tbp]
\centering\includegraphics[width=\linewidth]{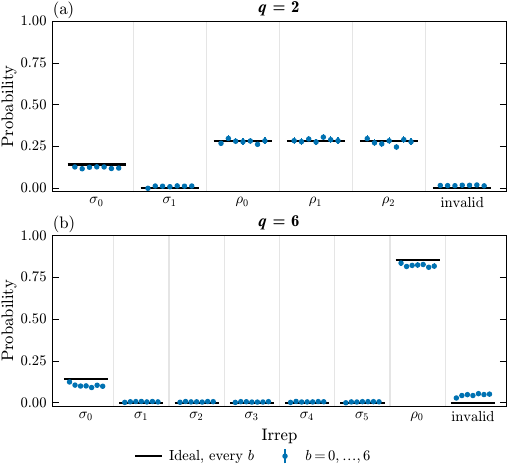}
\caption{Weak Fourier sampling for $p=7$ with (a) $q=2$ and (b) $q=6$, from the raw hardware data of the forgetful circuits (2,048 shots per secret, 95\% Wilson intervals). The irrep distribution is the same for every $b$, so weak Fourier sampling reveals nothing about the hidden subgroup. ``Invalid'' counts outcomes outside the group encoding.}

\label{fig:weak}
\end{figure}

First, in Fig.~\ref{fig:weak} we demonstrate that weak fourier sampling provides no information about the hidden subgroup parameterised by $b$. This draws a direct contrast with the abelian hidden subgroup problem where the irrep label revealed by weak fourier sampling uniquely determines the hidden subgroup. We draw our weak Fourier samples from the forgetful abelian circuits. On the $k=0$ branch, $F_q$ on the second register completes the non-abelian QFT on the one-dimensional irreps, so the measured $l$ is the label of $\sigma_l$; on the $k\neq0$ branch the irrep is fixed by $k$ alone, and weak sampling discards the second register anyway. For $k\neq0$ the irrep is $\rho_c$ with $c=(\log_g k-1)\bmod\frac{p-1}{q}$, where $g$ is the primitive root with $\alpha=g^{(p-1)/q}$, and for $k=0$ it is $\sigma_l$. For our subgroups $H_b$ only $l=0$ occurs, as seen in Eq.~\eqref{eq:finalqftstate}.

\begin{figure}[tbp]
\centering\includegraphics[width=\linewidth]{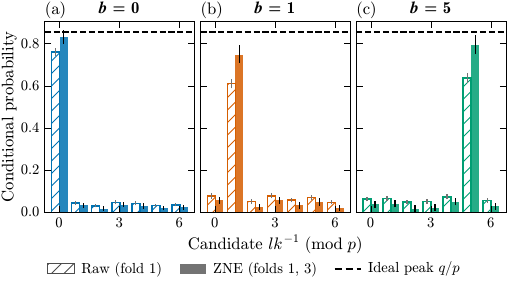}
\caption{Strong Fourier sampling in the adapted basis for $p=7$, $q=6$ with (a) $b=0$, (b) $b=1$ and (c) $b=5$ (2,048 shots per circuit; 95\% Wilson intervals for raw data and bootstrap intervals for ZNE). Conditioned on $k\neq0$, the candidate $lk^{-1} \bmod p$ peaks at the hidden $b$ with ideal height $q/p$.}

\label{fig:adapted}
\end{figure}

\begin{figure}[tbp]
\centering\includegraphics[width=\linewidth]{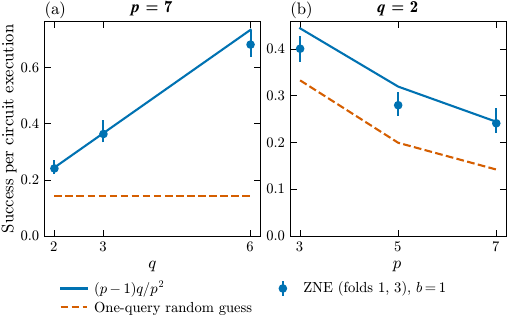}
\caption{Success probability of adapted-basis sampling for $b=1$, (a) against $q$ for $p=7$ and (b) against $p$ for $q=2$ (2,048 shots each, 95\% bootstrap intervals). The hardware follows the prediction $(p-1)q/p^2$, and for fixed $q$ it decays as $p$ increases.}

\label{fig:success}
\end{figure}

Fig.~\ref{fig:adapted} demonstrates that, unlike weak fourier sampling, strong fourier sampling in an adapted basis gives $b$ with high probability. Fig.~\ref{fig:success} shows that the success probability due to strong fourier sampling in the adapted basis follows the theoretical prediction of $(p-1)q/p^2$. Further, it shows that for fixed $q$ we have decaying success probability with increasing $p$, and hence we require a different sampling method for small $q$.

\begin{figure}[tbp]
\centering\includegraphics[width=\linewidth]{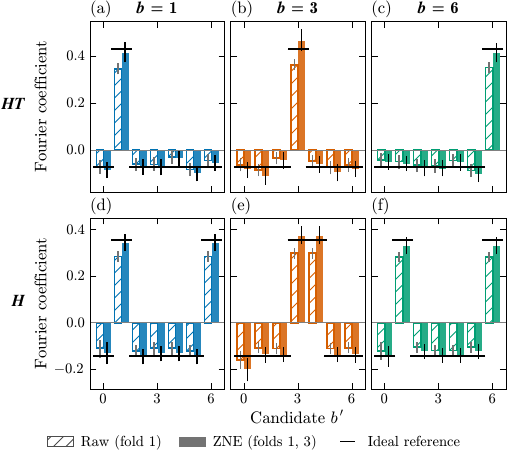}
\caption{Strong Fourier sampling of adjacent pairs for $p=7$, $q=2$ with $b=1,3,6$, applying (a)--(c) $HT$ and (d)--(f) $H$ to the least significant bit $t_0$ of $t$ (2,048 shots per circuit, 95\% bootstrap intervals). The Fourier coefficient at $b'$ is the real part of $e^{-i\pi/4}(-1)^t\omega_p^{-mb'}$ for $HT$, and of $(-1)^t\omega_p^{-mb'}$ for $H$, with $m=(1-\alpha)k \bmod p$, summed over the shots with $k\neq0$ and divided by the total number of shots. With $HT$ the signal peaks only at $b$, whereas $H$ cannot distinguish $b$ from $-b$.}

\label{fig:pairsfft}
\end{figure}

Strong sampling via adjacent pairs is demonstrated by Fig.~\ref{fig:pairsfft} and demonstrates a high probability of success even for small $q$. Fig.~\ref{fig:pairsfft} also confirms the theoretical prediction that the original protocol of MRRS for adjacent pairs fails to distinguish $b$ from $-b$. The theory predicts that we only need $O(\log p)$ samples to distinguish $b$. Unfortunately, we cannot confirm this prediction with our small sample of three different $p$. 

\begin{figure}[tbp]
\centering\includegraphics[width=\linewidth]{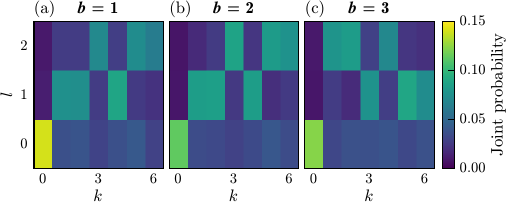}
\caption{Joint probability of $(k,l)$ under forgetful abelian sampling for $p=7$, $q=3$ with (a) $b=1$, (b) $b=2$ and (c) $b=3$, from raw hardware data (2,048 shots each). As $b=1$ and $b=2$ share the orbit $\{1,2,4\}$ they cannot be distinguished, whereas $b=3$ can.}

\label{fig:forgetful}
\end{figure}

The final example of sampling we consider is forgetful abelian sampling. Here the theory predicts that we will be able to distinguish the orbit of $b$ but not which $b$ we have within the orbit. Figure~\ref{fig:forgetful} demonstrates that this holds as we can successfully distinguish $b=1$ from $b=3$ but not from $b=2$ as $\{1,2,4\}$ share an orbit.

\begin{figure*}[tbp]
\centering\includegraphics[width=0.9\textwidth]{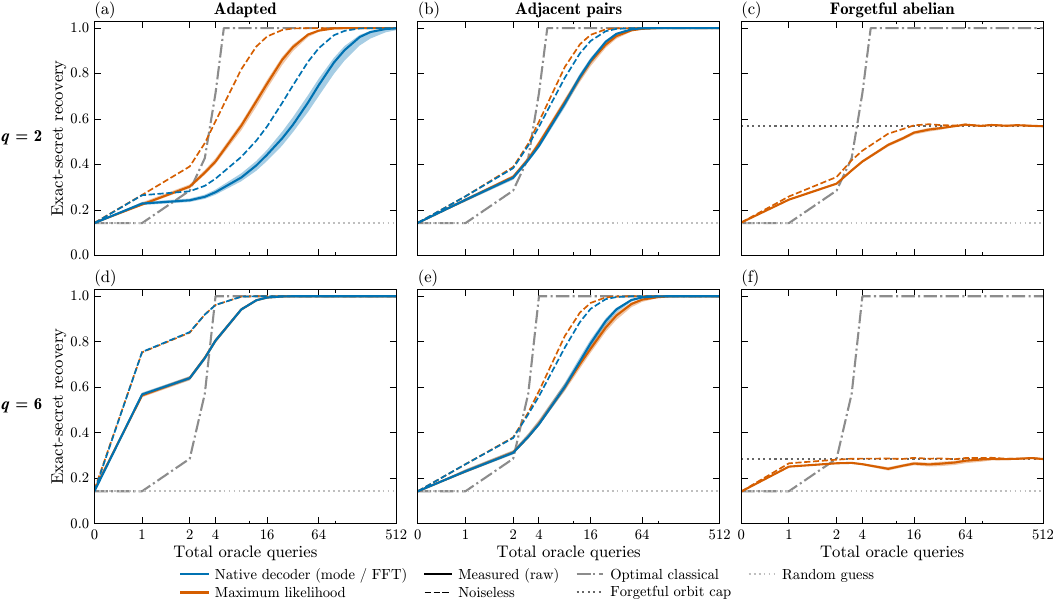}
\caption{Probability of recovering $b$ against total oracle queries for $p=7$ with $q=2$ (top) and $q=6$ (bottom), using (a,d) the adapted basis, (b,e) adjacent pairs and (c,f) forgetful abelian sampling. Each of 20,000 trials draws $b$ uniformly and resamples shots from that circuit's 2,048-shot histogram, one shot per query; bands are pointwise 95\% percentile intervals over 1,000 multinomial resamples of the histograms, so they include the hardware sampling error but not device drift, and maximum likelihood uses the ideal outcome probabilities. We compare the raw hardware data with noiseless simulation, a random guess and the optimal classical algorithm. Forgetful abelian sampling plateaus at $(p+q-1)/(pq)$.}

\label{fig:learning}
\end{figure*}

In Fig.~\ref{fig:learning} we compare the performance of each of the considered sampling techniques. To accomplish this we perform bootstrap sampling of the measured bitstrings (including those with $k=0$) and plot the probability of decoding a uniformly random $b$ successfully against the number of bootstrap samples taken (which is equivalent to the total oracle queries consumed). We do not use zero noise extrapolation for these plots as that would require additional queries. We include the recovery by random guess as a comparison that highlights the poor performance of weak fourier sampling. For a fair comparison we also include the optimum classical baseline of recovery probability against query count.

The standard classical approach to the HSP is the birthday algorithm \cite{simon1997power,nayak2021deterministic}. This proceeds by querying random elements until two return the same label, placing both elements in the same coset of $H$ which determines the secret via:
\begin{equation}
    b = (x'-x)(\alpha^t - \alpha^{t'})^{-1} \mod p
\end{equation}
where $f_b(x,t) = f_b(x',t')$. This requires $O(\sqrt{p})$ queries to succeed and is optimal up to a constant \cite{nayak2021deterministic,childs2017lecture}. However, choosing random elements to query is inefficient as some elements can never collide (e.g. if $t=t'$) and some different pairs test the same $b$. Therefore, we have chosen to use a derandomised version of the birthday algorithm \cite{nayak2021deterministic} as our classical comparison. In this derandomised version we have chosen the sequence of queries which ensures the maximum probability of successfully decoding $b$ for every query budget. We found this sequence via an exhaustive search and hence ensure this is the classical algorithm which uses the fewest possible queries. 

Despite the classical algorithm having the optimal success probability for all query counts, Fig.~\ref{fig:learning} shows that with up to 3 queries the real noisy quantum hardware gives a higher success probability for $p=7$ and $q=6$. Further, the noiseless simulation of quantum hardware exceeds 90\% success probability at fewer queries than the classical case. 

Fig.~\ref{fig:learning} provides a useful comparison between the relative performance of the different sampling methods at two extremes: $q=2$ (dihedral) and $q=6$ (affine). We see that as expected the adapted basis outperforms the adjacent pairs method for large $q$ but loses out to it for small $q$. It is also interesting to consider the behavior of the forgetful abelian sampling. We decode $b$ from the forgetful abelian signal using a maximum likelihood estimator. We see that the probability of successfully decoding $b$ via forgetful abelian sampling reaches a plateau before reaching $1$. This corresponds to the theoretical prediction that forgetful abelian sampling cannot distinguish $b$ within the same orbit, and as expected the success probability is capped at $\frac{p+q-1}{pq}$ in both figures. We can also see that the performance of the adjacent pairs sampling method is largely independent of $q$ as predicted by the theory, unlike the adapted basis method which is heavily dependent on $q$.

Fig.~\ref{fig:learning} also highlights the relative performance of our chosen decoders. For the adjacent-pairs method, FFT decoding performs similarly to maximum likelihood decoding for both $q=2$ and $q=6$. For the adapted-basis method, the mode decoder matches the maximum likelihood decoder at $q=6$, but significantly underperforms at $q=2$. This suggests that, for small $q$, the mode decoder does not exploit all the information available in the adapted-basis measurement outcomes.

\section{Discussion}
\label{section:discussion}
Importantly we have shown that the choice of measurement basis matters practically as well as asymptotically. Fig.~\ref{fig:learning} demonstrates that even at only $p=7$ choosing the weak Fourier or forgetful abelian methods prevents the success probability reaching $1$ irregardless of the number of samples taken. Even more starkly Fig.~\ref{fig:pairsfft} shows that a single gate can make the difference between being information-theoretic reconstructible and not. This highlights that choosing the correct measurement basis is crucial to ensure the problem is information-theoretic reconstructible as predicted by \cite{moore2007power}.

Figure~\ref{fig:learning} demonstrates that a system known to be fully reconstructible can outperform an optimized classical algorithm in recovery probability for low query counts despite the presence of hardware noise. This is despite the classical baseline having access to a perfect oracle, whereas every quantum query is subject to hardware noise. In noiseless simulation the separation extends further, with the quantum algorithm exceeding $90\%$ success at fewer queries than the classical algorithm.

In this work we have restricted ourselves to considering the q-hedral group, however, there are many other non-abelian groups which present interesting targets for near term quantum experiments. For many of these groups single-register measurements are provably insufficient, and the optimal measurement is entangled across several coset states~\cite{ettinger2004quantum,bacon2005optimal,hallgren2010limitations}, therefore a two-copy entangled measurement, or a single combination step of Kuperberg's sieve~\cite{kuperberg2005subexponential}, would be a natural next experiment. Whilst the dihedral group is included within the q-hedral groups we solve in this work it has more efficent solutions than the more general methods we have presented here. One recent algorithm \cite{lee2025learning} shows particularly strong promise for NISQ devices due to its constant circuit depth.

\paragraph*{Data and code availability} The experimental count histograms, quantum circuits, and analysis code supporting this work are available at \cite{o_brien_2026_23951178}. The repository includes the circuit-generation and compilation code, classical-baseline enumeration, decoding and uncertainty-analysis routines, and scripts used to generate the figures.

\paragraph*{Acknowledgements}
I thank Frank Verstraete for supporting the proposal to FormationQ for access to IonQ hardware and for encouraging me to publish this work independently. I thank Subhayan Roy Moulik for helpfully suggesting optimising the compilation of the quantum Fourier transform circuits. I am grateful to FormationQ and IonQ for providing access to the IonQ Forte trapped-ion quantum computer.

\paragraph*{Author contributions}
The author conceived and designed the study, carried out the experiments and analysis, and wrote the manuscript. The code used for circuit generation and compilation, classical-baseline enumeration, decoding, uncertainty analysis and figure generation was written with the assistance of Claude Opus 5.5. Claude was also used to generate the circuit diagrams in the Appendix and to review the manuscript before submission. The author checked and validated all code and results, and takes full responsibility for the content of this paper.

\bibliography{references}

\appendix
\section{Quantum Circuit}

\label{section:circuits}
\begin{figure*}[t]
\centering\includegraphics{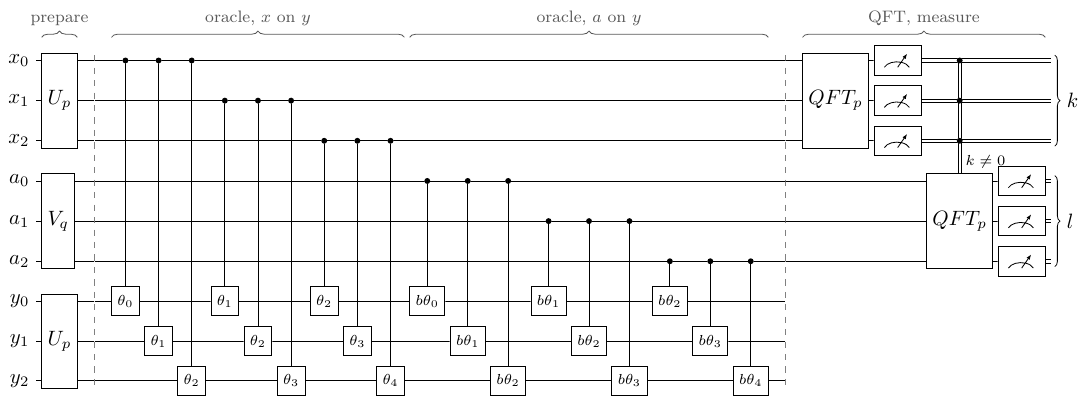}
\caption{Adapted-basis circuit for $p=7$ in the field encoding. Controlled phases $\theta_m=2\pi2^m/p$ from $x$ and $b\theta_m$ from $a$ act on the computation register $y$. $QFT_p$ on $x$ gives $k$ and $QFT_p$ on $a$ gives $l$; shots with $k=0$ are discarded. We omit the $y$ register's final $QFT_p^\dagger$ and $\pi$ as $y$ is discarded.}
\label{fig:circuit-adapted}
\end{figure*}
\subsection{Initial State}
In order to prepare the uniform superposition over $p$ computation basis states required by \eqref{eq:initialstate} we use the Shukla-Vedula algorithm \cite{shukla2024efficient} which uses $O(\log_2(p))$ qubits and $O(\log_2 (p))$ gates and we denote with $U_p$.

To prepare the state  $\frac{1}{\sqrt{q}}\sum_{t=0}^{q-1}\ket{\alpha^t}$ we employ the Gleinig-Hoefler algorithm \cite{10.1109/DAC18074.2021.9586240} via the open source PyEncode library \cite{suresh2026pyencode}. However, for large $q$ the state stops being sparse and this algorithm produces more 2 qubit gates than Qiskit's in-built dense state preparation technique \cite{iten2016quantum}, so we synthesise both and use whichever is cheaper. We denote this preparation technique with $V_q$.

\subsection{Oracle}
To implement the action of the oracle we use fourier conjugation. First, we prepare the computation register $y$ into a uniform superposition over $p$ basis states using the Shukla-Vedula algorithm ($U_p$):
\begin{equation}
   \ket{x}\ket{\alpha^t}\ket{0}\rightarrow\ket{x}\ket{\alpha^t}\ket{+_p}
\end{equation}
Then we act with $\lceil\log p\rceil^2$ controlled-phase gates with phase $\frac{2^{i+j}}{p}$ from the $i$-th qubit of the first register on the $j$-th qubit of the computation register:
\begin{equation}
    \ket{x}\ket{\alpha^t}\ket{+_p} \rightarrow \ket{x}\ket{\alpha^t}\frac{1}{\sqrt{p}}\sum_{y=0}^{p-1}\omega_p^{yx}\ket{y}
\end{equation}
where
\begin{equation}
    \omega_p^{yx} =  \prod_{i,j=0}^{n-1} e^{2\pi i \frac{2^{i+j}x_iy_j}{p}}
\end{equation}
where $x = \sum_i 2^i x_i$, $y = \sum_i 2^i y_i$. Similarly, we act with $\lceil\log p\rceil^2$ controlled-phase gates with phase $\frac{2^{i+j}}{p}b$ from the $i$-th qubit of the second register on the $j$-th qubit of the computation register to give overall transformation:
\begin{equation}
   \ket{x}\ket{\alpha^t}\ket{+_p}\rightarrow \ket{x}\ket{\alpha^t}\frac{1}{\sqrt{p}}\sum_{y=0}^{p-1}\omega_p^{y(x+\alpha^tb)}\ket{y}
\end{equation}
Next, we act with $QFT^{\dagger}_p$ upon the computation register which gives:
\begin{align}
    \ket{x}\ket{\alpha^t}\ket{+_p} &\rightarrow \ket{x}\ket{\alpha^t}\frac{1}{p}\sum_{x',y=0}^{p-1}\omega_p^{y(x+\alpha^t b-x')}\ket{x'}\\
    &\rightarrow \ket{x}\ket{\alpha^t}\ket{x+\alpha^tb}
\end{align}
Finally, we apply a random secret permutation $\pi$ to the computational basis elements of the computation register. Thus we have completed our application of the oracle:
\begin{equation}
   \ket{x}\ket{\alpha^t}\ket{0} \rightarrow \ket{x}\ket{\alpha^t}\ket{\pi(x+\alpha^tb)}
\end{equation}
As we never use the outcome of the measurement on the computation register in our algorithm we are free to instead discard the register. This would produce the coset state:
\begin{equation}
    \frac{1}{qp}\sum_{c=0}^{p-1}\sum_{t,t'=0}^{q-1} \ket{\pi^{-1}(c) - \alpha^tb}\bra{\pi^{-1}(c) - \alpha^{t'}b} \otimes \ket{\alpha^t}\bra{\alpha^{t'}}
\end{equation}
Under this assumption we are free to apply any unitary we wish to the discarded register before we discard it, and hence we do not actually need to implement the $QFT_p$ or permtuation $\pi$ as physical circuits provided we discard the measurement outcome obtained from the computation register.

\subsection{Non-Abelian QFT}
Once we have applied the oracle we must perform $QFT$ over the non-abelian q-hedral group. By inspecting \eqref{eq:finalqftstate} it is clear how to perform the non-abelian QFT as a quantum circuit. First, one must apply the abelian $QFT_p$ on the first register ($QFT_p \otimes \mathbb{I}_q$) giving state:
\begin{equation}
    \frac{1}{\sqrt{pq}} \sum_{k=0}^{p-1}\sum_{t = 0}^{q-1} \omega_p^{k (\pi^{-1}(c) - \alpha^tb)}\ket{k}\ket{\alpha^t}
\end{equation}
This can be rewritten as:
\begin{equation}
  \frac{1}{\sqrt{pq}}\sum_{t=0}^{q-1} \ket{0}\ket{\alpha^t}+ \frac{1}{\sqrt{pq}} \sum_{k=1}^{p-1}\sum_{t = 0}^{q-1} \omega_p^{k (\pi^{-1}(c) - \alpha^tb)}\ket{k}\ket{\alpha^t}
\end{equation}
Therefore, if we uncompute $\frac{1}{\sqrt{q}}\sum_{t=0}^{q-1}\ket{\alpha^t}$ controlled upon the first register being in state $\ket{0}$ then we apply the full non-abelian QFT giving state:
\begin{equation}
  \frac{1}{\sqrt{p}} \ket{0}\ket{0} +\frac{1}{\sqrt{pq}} \sum_{k=1}^{p-1}\sum_{t = 0}^{q-1} \omega_p^{k (\pi^{-1}(c) - \alpha^tb)}\ket{k}\ket{\alpha^t}
\end{equation}
This uncomputation is simply the inverse of the state preparation circuit ($V_q^{\dagger}$) we used to prepare $\frac{1}{\sqrt{q}}\sum_{t=0}^{q-1}\ket{\alpha^t}$. As we discard the results of the second register if the first register is measured to be $k=0$, we do not actually need to perform this uncomputation step, and hence we do not actually demonstrate non-abelian QFT, but rather abelian QFT with post-selection. 

The full adapted-basis circuit, combining the state preparation, oracle and QFT described above, is shown in Fig.~\ref{fig:circuit-adapted}.

\subsection{Quantum Fourier Transform}

The quantum Fourier transform over $\mathbb{Z}_p$
is a unitary on $\mathrm{span}\{\ket{j}\}_{j<p}$ acting as
\begin{equation}
    QFT_p \ket{j} = \frac{1}{\sqrt{p}}\sum_{k=0}^{p-1} \omega_p^{jk} \ket{k} \qquad \forall \quad  j \in \mathbb{Z}_p
\end{equation}
realised on $n=\lceil\log_2 p\rceil$ qubits. Its action on the remaining $2^n-p$ dimensions is unconstrained.

There exist many asymptotically efficient methods for implementing $QFT_p$ for generic $p$ \cite{kitaev1995quantum,hales2000improved, hales2002quantum, mosca2004exact, kuo2026quantum}. However, for small $p$ it is possible to find more hardware efficient implementations through numerical
synthesis, in the manner of approximate quantum compiling \cite{khatri2019quantum, madden2022best, rakyta2022approaching, nemkov2023efficient}. In this work we use an ansatz with alternating layers of arbitrary single-qubit gates and a round-robin of 2-qubit $R_{zz}$ gates. Every $QFT_p$ in our circuits is directly followed by a full $Z$ basis measurement, so the phase of each output is irrelevant. Hence, for a given ansatz $V(\Theta)$ we use BFGS to maximise $\mathcal{F}(\Theta)$:
\begin{align}
    \mathcal{F}(\Theta)
 &= \max_{\{\varphi\}}\left| \frac{1}{p}\sum_{k<p}
    \bra{k}V(\Theta) QFT_p^{\dagger}\ket{k}e^{-i\varphi_k }\right|\\
 &= \frac{1}{p}\sum_{k<p}\left|\langle k|V(\Theta) QFT_p^{\dagger}|k\rangle\right| 
\end{align}
The solution is then further refined using Gauss–Newton. We increment the depth of the ansatz until we achieve a machine precision approximation. Using this method we are able to attain far lower 2-qubit gate counts than that obtained by a standard Qiskit method of quantum Shannon decomposition of the embedded unitary $QFT_p \oplus \mathbb{I}_{2^{n}-p}$ as shown in Table~\ref{tab:fourier}.

\begin{table}[t]
\caption{Two-qubit ($R_{ZZ}$) gate counts for $QFT_d$ on $n=\lceil\log_2 d\rceil$ qubits, comparing our numerically synthesised circuits with Qiskit's quantum Shannon decomposition of $QFT_d \oplus I_{2^n-d}$.}

\label{tab:fourier}
\begin{tabular}{lccc}
\hline
Transform & Qubits & Ours & Qiskit \\
\hline
$\mathrm{QFT}_3$ & 2 & 2  & 3  \\
$\mathrm{QFT}_5$ & 3 & 8  & 19 \\
$\mathrm{QFT}_6$ & 3 & 7  & 19 \\
$\mathrm{QFT}_7$ & 3 & 10 & 19 \\
\hline
\end{tabular}
\end{table}

\subsection{Exponent encoding}

\begin{figure}[t]
\centering
\includegraphics[width=0.5\linewidth]{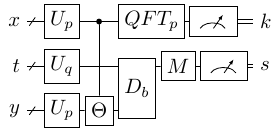}\par \medskip
\includegraphics[width=\linewidth]{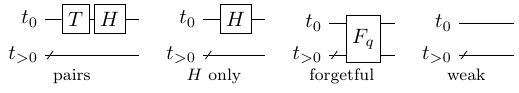}
\caption{Exponent-encoding circuit. $\Theta$ applies the controlled phases from $x$ on the computation register $y$ as in Fig. 7, and $D_b$ applies the $b$-dependent phases from $t$ (Fig. 9). The measurement $M$ on $t$ sets the sampling method: $HT$ on $t_0$ for adjacent pairs, $H$ for the original MRRS scheme, $F_q$ for forgetful abelian sampling and no gate for weak Fourier sampling.}

\label{fig:circuit-exponent}
\end{figure}
In Sections~\ref{section:pairs} and \ref{section:forgetful} we use an exponent encoding of our group rather than the above consider field encoding, with the full circuit shown in Fig.~\ref{fig:circuit-exponent}. This greatly simplifies the state preparation of our circuit which can now use the Shukla-Vedula algorithm \cite{shukla2024efficient} for both registers, which we denote $U_p$ and $U_q$, to produce:
\begin{equation}
    \frac{1}{\sqrt{qp}}\sum_{x = 0}^{p-1}\sum_{t=0}^{q-1} \ket{x}\ket{t}
\end{equation}
However, it does increase the complexity of applying the oracle as we can no longer use $n^2$ controlled-phase gates to apply the contribution from the second register. Instead we act with $n$ different $n_q$-qubit diagonal gates, which we collectively denote $D_b$, acting on $\ket{t}$ and the $i$-th qubit of the $\ket{y}$ register ($n_q =\lceil \log q \rceil + 1$):
\begin{equation}
    \sum_{y_i=0}^1\sum_{t=0}^{q-1} e^{2\pi i 2^i\frac{y_i \alpha^t b}{p}}\ket{t}\bra{t} \otimes \ket{y_i}\bra{y_i}
    \label{eq:diagonalgateorigingal}
\end{equation}

We synthesize these gates using techniques introduced by \cite{Welch_2014}. We first rewrite $f(t)= 2\pi  \frac{2^i\alpha^t b}{p}$ as a Walsh-Fourier series given by $f(t) = \sum_{s=0}^{2^{n_q}-1} \hat f(s) (-1)^{\sum_{j}s_jt_j}$. This reduces \eqref{eq:diagonalgateorigingal} to:
\begin{equation}
        \sum_{y_i=0}^1\sum_{t=0}^{q-1} \prod_{s=0}^{2^{n_q}-1}
        e^{iy_i \hat f(s)(-1)^{\sum_{j}s_jt_j}}\ket{t}\bra{t} \otimes \ket{y_i}\bra{y_i}
\end{equation}
We define the Walsh operator $W_s$ as:
\begin{equation}
    W_s = \bigotimes_{j=1}^{n_q} (Z_j)^{s_j} = (Z_1)^{s_1}\otimes (Z_2)^{s_2} \otimes \cdots \otimes (Z_{n_q})^{s_{n_q}}
\end{equation}
Hence we can further reduce \eqref{eq:diagonalgateorigingal} to:
\begin{equation}
        \sum_{y_i=0}^1\sum_{t=0}^{q-1} \prod_{s=0}^{2^{n_q}-1}
        e^{i\hat f(s)\frac{1-Z_{y_i}}{2} W_s}\ket{t}\bra{t} \otimes \ket{y_i}\bra{y_i}
\end{equation}
We implement this gate by acting with CNOTs upon the register $t$ such that every parity $\oplus_{j}t_j^{s_j}$ occurs at least once on some qubit. When that parity is stored we can then perform $e^{-i\frac{\hat f(s)}{2} Z_{y_i}W_s}$ with a single $R_{zz}$ between the parity qubit and the $i$-th qubit of $\ket{y}$ and $e^{i\frac{\hat f(s)}{2} W_s}$ with a single $R_z$ gate just on the parity qubit. 

\begin{figure}[t]
\centering
\includegraphics{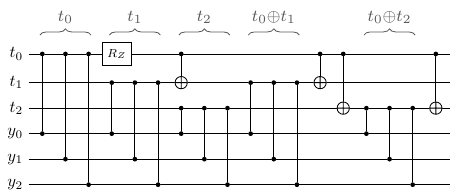}
\caption{Walsh-series implementation of $D_b$ for $p=7$, $q=6$. A CNOT walk stores each required parity of $t$ on a single qubit, which is then coupled to each qubit of $y$ with an $R_{ZZ}$ gate.}

\label{fig:oracle-diagonal}
\end{figure}

As $f(t)$ is only restricted on $t \in \mathbb{Z}_q$ we need only have $q$ non-zero $\hat f(s)$. Therefore, we do not need to implement $W_s$ for every possible parity. Hence, we exhaustively search all the different methods of iterating through $q$ parities to find the shortest CNOT sequence that we can use to implement this gate. Figure~\ref{fig:oracle-diagonal} shows an example of our implementation of $D_b$ for $p=7, q=6$.

This methodology results in fewer 2-qubit gates for $D_b$ than using the standard Qiskit synthesis for diagonal gates \cite{shende2005synthesis} as highlighted in Table~\ref{tab:diagonal}.
\begin{table}[t]
\caption{Two-qubit ($R_{ZZ}$) gate counts for the oracle diagonal $D_b$ at $p=7$, comparing our Walsh-series synthesis with one Qiskit diagonal gate per qubit of $y$. Counts are the same for every $b$.}
\label{tab:diagonal}
\begin{tabular}{cccc}
\hline
$q$ & Qubits ($t$+$y$) & Ours & Qiskit \\
\hline
2 & 1+3 & 3  & 3  \\
3 & 2+3 & 6  & 15 \\
6 & 3+3 & 19 & 39 \\
\hline
\end{tabular}
\end{table}

\subsection{Zero-noise extrapolation}
\begin{figure}[htbp]
\centering\includegraphics{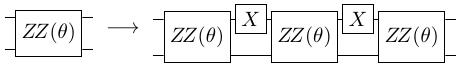}
\caption{Gate folding for zero-noise extrapolation. Each $ZZ(\theta)$ is replaced by $ZZ(\theta)ZZ(\theta)^{-1}ZZ(\theta)$ with $ZZ(\theta)^{-1}=X_1\,ZZ(\theta)\,X_1$ which triples the two-qubit gate count. Single-qubit gates are not folded, and circuits are submitted as native gates so the folds are not compiled away.}

\label{fig:circuit-fold}
\end{figure}

Zero-noise extrapolation (ZNE) \cite{li2017efficient, temme2017error} runs a circuit at several deliberately amplified noise levels and extrapolates the measured quantities back to zero noise. We perform digital zero-noise extrapolation through unitary folding \cite{giurgica2020digital} and exponential extrapolation \cite{endo2018practical}. We deliberately increase the noise in our circuit by replacing every $R_{zz}(\theta)$ gate by $R_{zz}(\theta)R_{zz}(\theta)^{-1}R_{zz}(\theta)$ as shown in Fig.~\ref{fig:circuit-fold}. We chose to just scale the noise due to our 2-qubit gates as they are the dominant contribution to error in our circuits. We assign the folded circuit a noise scale of $\lambda=3$ and the original circuit $\lambda = 1$. Assuming the noise drives the state towards the maximally mixed state, so that the signal decays exponentially with $\lambda$, we model the expectation values as
\begin{equation}
    \braket{O_{\lambda}} - \frac{\operatorname{Tr}[O]}{d} = g^{\lambda}\left(\braket{O_0} - \frac{\operatorname{Tr}[O]}{d}\right)
\end{equation}
where $d$ is the Hilbert-space dimension, so that $\operatorname{Tr}[O]/d$ is the expectation value of $O$ in the maximally mixed state. Hence, we can derive:
\begin{equation}
    \braket{O_0} = \frac{\operatorname{Tr}[O]}{d} + \frac{\braket{O_1} - \frac{\operatorname{Tr}[O]}{d}}{g}
\end{equation}
We fit a single $g$ per circuit by least squares across all outcomes.

To derive error bars for our ZNE values we perform bootstrap resampling from the measured count histograms. For each replicate we resample both the $\lambda=1$ and $\lambda=3$ histograms, refit $g$ and recompute the ZNE value. We perform 1000 replicates and record the 2.5th and 97.5th percentiles for the ZNE value. 

\end{document}